\documentstyle[psfig]{basi}
\begin{document}
\title[Spectroscopic Survey of Field stars: A search for Metal-Poor stars] 
      {Spectroscopic Survey of Field stars: A search for Metal-Poor stars} 
\author[Sunetra Giridhar and Aruna Goswami]
       {Sunetra Giridhar \thanks{e-mail : giridhar@iiap.ernet.in}
 and  Aruna Goswami\\ 
        Indian Institute of Astrophysics, Bangalore 560 034}

\maketitle

\begin{abstract}
 We have undertaken a spectroscopic survey  of field stars to find    
 metal-poor objects  among them. Though the main objective of the survey is
 to find new metal-poor stars, stellar parameterization is carried out
 for all the sample stars so that the other categories of interesting objects 
 like composite stars, weak or strong CN, CH stars etc. can also
  be identified. 

     Observations are carried out using OMR spectrograph attached to 
  VBT, Kavalur. The sample of candidate stars are chosen from prismatic
  survey  of Beers and his collaborators covering a large part of the 
  Galaxy. At the first phase of this project, the analysis is completed
  for a set of 19 relatively hot stars (T$_{eff}$ in 6000 to
  8000K range).   The metallicities of the program stars 
  are derived by synthesizing the spectrum in the wavelength range 4900 to
  5400 $\AA$~ for different metallicities and matching them with the observed
  spectra.
 This spectral region  contains strong feature of FeI at 5269 $\AA$~ 
 and one moderately strong Fe I blend at 5228 $\AA$. These features were 
 generally relied upon for Fe/H determination. 
 More than half of the candidate stars were found to show [Fe/H] in
 -0.7 to -1.2 range. Two most metal-poor stars have [Fe/H] values of -1.3
 and -1.8. It appears that metal-poor candidates suggested by Beers et al.
 from their prismatic survey has a very significant fraction of metal-poor
 stars. The significantly metal-poor stars
  found so far would be studied in detail using high resolution spectra
  to understand nucleosynthesis processes that might have occurred in
  early Galaxy.  
\end{abstract}

\section{Introduction}
 Spectroscopic surveys containing large areas of the sky provide wealth of
 information on different groups of stars.  Although  finding new 
  metal-poor stars is the   prime objective of many surveys, many interesting objects
 like emission line objects, stars with unusually strong or weak CH, CN
  bands etc. are also discovered. Discovery and comprehensive studies
 of metal-poor stars in different Galactic locations are very important
 tools to study the structure and evolution of early Galaxy.

 In the  past, the surveys relied upon kinematic data for more
 efficient metal-poor star detection and hence the objects were
 chosen from  the proper-motion catalogues. It was pointed out by Carney 
 (1997) and others that  for complete understanding of disk-halo structure, the
  kinematic bias should be avoided.

                                Objective prism survey of Beers, Preston and
 Shectman (1992) called HK survey covers large portion of the sky in northern
 and southern hemispheres. Hamburg/ESO survey covers a much larger
 portion of the sky though it is restricted to the southern hemisphere.
 Edinburgh -Cape blue object survey of Stobie et al. 1997
( hereinafter EC)  covers most of the high galactic
 latitude southern sky objects. 

 In our earlier work (Giridhar, et al. 2001), we have carried out spectroscopic
 investigation of stars  selected from the HK survey and from a list
 of stars with tangential velocity in excess of 100 kms$^{-1}$ published
 by Lee (1984). We found the stars from HK survey to be considerably
 metal-poor, but the stars taken from Lee's table were not very promising
 metal-poor candidates.
\section{Selection of Program stars}
  The program stars listed in Table 1, are mostly  taken from the EC         
survey. 
\begin{table}
{\bf Table 1: EC stars and standards observed with VBT }\\
\begin{tabular}{cccccc}
             &       &         &         &      &   \\
\hline
{\bf Star}   &{\bf V } & {\it l} & {\it b} &   RA (2000)    & DEC (2000)     \\

\hline
EC 05148-2731 & 12.69 &  229.9  & -31.8   & 05:16:50.7  &  -27:28:25  \\
EC 10488-1244 & 11.15 &  263.1  &  40.4   & 10:51:18.6   &  -13:00:28  \\ 
EC 10004-1405 & 13.65 &  253.1  &  31.8   & 10:02:54.5  & -14:19:41    \\
EC 10292-0956 & 12.06 &  255.9  &  39.6   & 10:31:44.9 & -10:11:36    \\ 
EC 11091-3239 & 12.56 &  279.8  &  25.4   & 11:11:32.6  & -32:56:17  \\ 
EC 11175-3214 & 12.55 &  281.5  &  26.5   & 11:19:58.8  &  -32:30:34   \\
EC 11553-2731 & 11.29 &  288.7  &  33.6   & 11:57:55.3  & -27:48:22    \\
EC 11260-2413 & 12.85 &  279.9  &  34.6   & 11:28:29.8  & -24:29:38    \\
EC 12245-2211 & 13.96 &  295.6  &  40.1   & 12:27:07.8 & -22:27:44   \\
EC 12418-3240 & 12.19 &  301.3  &  29.9   & 12:44:32.6 & -32:57:15    \\
HD 86801      &  8.79 &  200.5  &  52.6   & 10:01:34.9 & +28:34:00 \\
HR 5694       &  5.10 &    3.7  &  46.4   &  15:19:18.7 & +01:45:55.4 \\
EC 12473-1945 & 10.43 &  302.5  &  42.8   & 12:49:57.0  & -20:02:04    \\
EC 12473-1711 & 11.68 &  302.4  &  45.4   & 12:49:58.3  & 17:27:44    \\
EC 12477-1724 & 11.40 &  302.6  &  45.2   & 12:50:24.1  & -17:40:20    \\
EC 12493-2149 & 13.51 &  303.1  &  40.8   & 12:51:58.6 & -22:06:09    \\
EC 13042-2740 & 12.95 &  307.1  &  34.8   & 13:06:55.8  & -27:56:29    \\
EC 13501-1758 & 12.63 &  322.8  &  42.3   &  13:52:50.7 & -18:13:21    \\
HR 5384       &  6.30 &  347.2  &  56.0   & 14:23:15.2  & +01:14:29.6  \\
EC 13506-1845 & 12.87 &  322.6  &  41.5   & 13:53:21.7  & -19:00:32    \\
EC 13564-2249 & 12.78 &  322.6  &  37.2   &  13:59:15.3 &  -23:03:36   \\
EC 13567-2235 & 12.60 &  322.8  &  37.4   &  13:59:29.8 & -22:49:57    \\
EC 13586-2220 & 13.72 &  323.4  &  37.5   & 14:01:27.4 & -22:34:41    \\
EC 14005-2224 & 13.33 &  323.9  &  37.3   & 14:03:22.8  &  -22:38:50   \\
EC 15473-1241 & 13.11 &  356.1  &  31.1   & 15:50:06.8  & -12:50:11    \\
HR 4054       &  4.80 &  217.1  &  54.4   & 10:19:44.1 & +19:28:15 \\
\hline
\end{tabular}
\end{table}  
 
EC survey  was initiated to study 
distant hot stars  from their photographic photometry.
  While picking hot stars in B versus (U-B) diagram, some F and G stars 
which are in fact metal-poor were mistaken as blue objects. The U colours had
  become brighter due to decreased absorption from otherwise strong metal lines.
  A shorter list (of nearly 600 stars) of  such contaminating
 (but interesting) objects was supplied
  by Beers et al. (2001) from which stars in the magnitude range 11-14 mag  
 were observed at VBT using OMR spectrograph. In addition to these objects, 
spectrophotometric and radial velocity standards were also observed.
 In this paper, we present the results  obtained at the first phase of 
  an ongoing spectroscopic survey program.

  Broadband UBV colours of  the EC stars are available in Beers et al. (2001).
 The errors in this photoelectric photometry are in the range of 0.02 to 0.04.
  These authors have estimated reddening by interpolation in  the tables of
  Burstein \& Heiles (1982). The reddening estimates  have uncertainties
  of $\sim$ $\pm$ 0.03. 
\section{Observation and Data analysis}
The spectrograph used for observation is the OMR Spectrograph designed and built
by the Optomechanics Research  Inc., Vail, Arizona, USA. It is available
 at the Cassegrain focus of the 2.3m Vainu Bappu Telescope ( VBT) at 
Kavalur (Prabhu, Anupama \& Surendiranath 1998). 
  With a 600 lmm$^{-1}$ grating, we get a dispersion of 2.6$\AA$~ per 
pixel.  The detector is Tetronix 1K CCD with 24$\mu$m pixels. 
 Wavelength calibration was done with the help of Fe-Ar hollow cathode lamp.
  Flat field correction was done using  Tungsten-Halogen Quartz   lamp.
The spectroscopic reduction steps were carried out using various tasks
 included in the NOAO package  of IRAF.          
 At a resolution of 1000, one can  only   measure the strengths of 
 blends around  strong features like hydrogen lines, Ca II H, K lines etc.
  Although we could identify a few high radial velocity objects, 
  good radial velocity measurements using the cross-correlation technique 
  could not be made as the radial velocity standards observed with the same
  setup were not of the same spectral types as program stars for some 
  nights. We, therefore, do not report the radial velocities though  good
  measurements exist for one third of the sample.

\section{Determination of the stellar parameters}
    Firstly, we have attempted a preliminary classification of these unknown 
objects.   Towards this objective, we were guided by a 
    library of stellar spectra published by Jacoby, Hunter and Christian 1984. 
   This library contains  161 flux calibrated spectra of O-M stars covering
   luminosity class I to V. Most of these stars are of  luminosity types
   I, III and V with a very few stars of type II. 

   We have chosen  this collection of standard stars primarily because, these 
spectra are observed with nearly the same resolution as the spectra obtained by
 us at VBT. Catalogue stars used for our purpose are listed in Table 2.
\begin{table}
{\bf Table 2: Line strengths in \AA~  measured for the Catalogue stars }\\
\begin{tabular}{cccccccccccc}
         &      &     &     &      &      &     &     &     &     &     &    \\
\hline
         &      &     & Ca II K & Ca II H & $H_{\delta}$ & Ca I & $H_{\gamma}$& $H_{\beta}$&Mg+Fe & Mg+Fe & Fe I  \\
{\bf Star }&$(B-V)_{o}$& {\bf Sp type}& 3934& 3969& 4102& 4227& 4340& 4861& 5171& 5184& 5270 \\ 
\hline
HD 9547   & 0.30 & A5V & --- &  --- & 12.8 & 0.7 &12.4 &13.8 & 1.3 & 0.4 & 1.5    \\
HD 10032  & 0.33 & F0V & 6.0 &  9.2 &  7.3 & 0.9 & 6.2 & 8.0 & 1.5 & 0.7 & 1.7    \\ 
HD 23733  & 0.34 & A9V & 6.6 & 10.1 &  8.3 & 0.8 & 7.2 & 8.4 & 1.3 & 0.6 & 2.3    \\ 
SAO 57199 & 0.40 & F6V & 9.9 & 10.3 &  4.1 & 1.1 & 4.6 & 4.6 & 2.0 & 1.0 & 1.1    \\
HZ 948    & 0.47 & F3V & 7.4 &  7.8 &  5.3 & 0.8 & 5.3 & 4.9 & 1.8 & 0.6 & 2.0    \\  
HD 5702   & 0.49 & F7V & 10.8& 10.2 &  4.0 & 1.5 & 3.9 & 4.8 & 2.1 & 1.1 & 2.0    \\ 
HD 107132 & 0.50 & F7V & 11.3& 10.8 &  4.3 & 1.2 & 3.3 & 3.8 & 2.0 & 1.0 & 2.7    \\ 
HZ 227    & 0.52 & F5V & 8.9 & 8.93 &  5.7 & 0.7 & 5.0 & 6.8 & 1.9 & 1.1 & 1.2    \\ 
HR 5694   & 0.54 & F8IV& 10.2&  9.3 &  4.9 & 0.9 & 3.4 & 4.8 & 1.9 & 0.8 & ---    \\
HD 6111   & 0.55 & F8V & 12.1& 10.9 &  3.3 & 1.4 & 3.5 & 3.4 & 2.4 & 1.4 & ---    \\ 
HD 107399 & 0.56 & F9V & 13.3& 12.2 &  2.7 & 1.6 & 2.9 & 3.7 & 2.4 & 1.3 & 2.7    \\
HD 31084  & 0.57 & F9V & 13.0&  9.2 &  3.7 & 1.3 & 3.1 & 3.8 & 2.5 & 1.2 & ---    \\ 
HD 107214 & 0.58 & F7V & 11.0&  9.2 &  3.5 & 1.5 & 3.9 & 4.0 & 2.3 & 1.3 & ---    \\
HD 66171  & 0.60 & G2V & 13.1& 11.0 &  2.3 & 1.9 & 2.6 & 2.6 & 3.1 & 1.6 & ---    \\ 
HD 17647  & 0.61 & G1V & 13.9& 10.5 &  2.5 & 2.3 & 3.0 & 2.8 & 3.4 & 2.0 & 3.7    \\
HR 5384   & 0.63 & G1IV& 12.9& 10.0 &  3.3 & 1.4 & 3.0 & 3.5 & 3.0 & 1.7 & ---    \\
BD 58199  & 0.64 & G3V & 14.8& 11.6 &  2.1 & 1.9 & 3.8 & 2.5 & 3.1 & 2.0 & 2.5    \\ 
HD 28099  & 0.66 & G0V & 8.7 & 12.2 &  5.8 & 1.3 & 4.0 & 3.6 & 3.3 & 1.7 & 4.0    \\
HD 22193  & 0.71 & G6V & 13.6& 10.0 &  2.3 & 2.3 & 3.0 & 2.6 & 3.8 & 2.0 & 3.7    \\
TrA 14    & 0.74 & G4V & 14.0& 12.0 &  2.1 & 1.2 & 2.0 & 1.7 & 1.8 & 1.5 & 2.6    \\
HD 12027  & 0.22 &A3III& 3.3 &  --- & 13.8 & 0.4 &16.3 &18.4 & 1.3 & 0.3 & ---    \\   
HD 12161  & 0.23 &A8III& 4.8 & 11.3 & 12.5 & 0.6 & 8.1 &11.9 & 1.6 & 0.5 & 0.8    \\ 
HD 240296 & 0.25 &A6III& 5.6 &  8.2 &  9.3 & 1.2 & 9.0 & 9.0 & 1.7 & 0.4 & 1.7    \\
HD 64191  & 0.28 &F0III& 6.6 & 10.5 & 10.8 & 0.6 & 9.9 & 7.1 & 1.4 & 0.7 & 1.3    \\
BD 003227 & 0.40 & F5II& 8.7 &  9.2 &  4.8 & 0.8 & 4.2 & 4.6 & 1.8 & 1.0 & 2.0    \\ 
HD 56030  & 0.45 &F6III& 8.9 &  9.2 &  3.9 & 1.3 & 3.6 & 4.5 & 2.4 & 0.8 & 1.5    \\
BD 610367 & 0.46 &F5III& 8.7 &  9.4 &  5.9 & 1.5 & 5.2 & 5.8 & 1.4 & 0.7 & 2.8    \\  
HD 5211   & 0.47 &F4III& 10.0&  9.9 &  5.1 & 1.1 & 4.5 & 5.0 & 1.8 & 0.8 & 1.3    \\ 
SAO 20603 & 0.48 &F7III& 11.4&  9.9 &  3.7 & 1.2 & 3.9 & 4.4 & 2.3 & 1.0 & 2.8    \\ 
HD 9979   & 0.50 &F8III& 14.1& 12.2 &  3.4 & 2.3 & 3.0 & 3.1 & 3.5 & 2.0 & 1.9    \\ 
BD 302347 & 0.58 &G0III& 13.6& 10.5 &  3.5 & 1.6 & 3.3 & 3.8 & 2.6 & 1.2 & 2.8    \\
HD 15866  & 0.66 &G0III& 12.5& 11.8 &  3.0 & 1.5 & 2.8 & 4.0 & 2.6 & 1.4 & 3.4    \\
HD 25894  & 0.71 &G2III& 13.1& 10.4 &  3.0 & 2.1 & 2.6 & 2.9 & 3.8 & 2.0 & 4.1    \\
HD 2506   & 0.89 &G4III& 18.6& 13.0 &  1.9 & 2.3 & 1.9 & 2.0 & 2.8 & 1.2 & 3.0    \\
HR 4786   & 0.89 &G5III& 17.2& 12.3 &  2.2 & 1.2 & 2.0 & 3.1 & 3.1 & 1.1 & ---    \\ 
BD 281885 & 0.91 &G5III& 17.7& 15.0 &  1.7 & 3.0 & 2.6 & 2.5 & 3.5 & 1.8 & 4.2    \\  
HD 249384 & 1.03 & G8II& 18.0& 16.6 &  1.8 & 2.9 & 2.4 & 1.4 & 4.6 & 2.9 & 4.0    \\
HD 250368 & 1.04 & G8II& 14.0& 12.0 &  1.6 & 1.9 & 1.8 & 1.4 & 3.9 & 2.0 & 4.0    \\
SAO 12149 & 0.07 & A1I & 1.4 &  5.4 &  4.7 & 0.4 & 4.4 & 4.8 & 0.9 & 0.1 & 0.7    \\
SAO 12096 & 0.13 & A4I & 5.4 &  9.5 &  7.3 & 0.7 & 7.3 & 8.2 & 1.2 & 0.4 & ---   \\
HD 842    & 0.14 & A9I & 6.6 & 10.1 &  8.9 & 0.8 & 7.9 &10.0 & 1.5 & 0.6 & 2.2    \\
SAO 37370 & 0.16 & F0Ib& 6.7 &  9.9 &  7.1 & 0.6 & 6.3 & 7.5 & 1.3 & 0.6 & 2.3    \\
42LSI    & 0.17 & A2I & 6.5 &  7.7 &  9.1 & 0.7 & 7.2 & 7.8 & 1.5 & 0.5 & 2.2    \\
HD 9167   & 0.18 & A7I & 6.5 &  8.6 &  6.1 & 0.7 & 5.9 & 5.6 & 1.4 & 0.4 & 2.6    \\
BD 580204 & 0.21 & F2I & 8.5 & 10.9 &  7.1 & 0.7 & 7.6 & 7.8 & 1.3 & 0.6 & 2.8    \\
SAO 21536 & 0.28 & F4I & 9.1 & 11.8 &  5.7 & 0.9 & 6.3 & 6.9 & 2.0 & --- & 3.1    \\
HD 9973   & 0.30 &F5Iab&13.0 & 12.7 &  5.4 & 1.9 & 5.7 & 5.2 & 2.7 & 1.2 & 4.4    \\
HD 8992   & 0.31 & F6Ib&13.0 & 14.9 &  4.0 & 2.0 & 4.9 & 4.3 & 2.2 & 0.8 & ---    \\
HD 187428 & 0.56 & F8Ib&14.0 & 13.47&  3.2 & 1.3 & 4.3 & 4.1 & 2.4 & 0.9 & 3.8    \\
HD 17971N & 0.60 & F7I &12.9 & 12.0 &  4.3 & 1.4 & 4.9 & 6.1 & 2.9 & 1.2 & 3.2    \\
HD 25361  & 0.88 & G0Ia&15.5 & 13.2 &  2.4 & 1.8 & 3.5 & 3.8 & 2.5 & 1.6 & 4.7    \\
SAO 21446 & 0.91 & G1I &18.4 & 14.0 &  2.1 & 2.2 & 3.3 & 3.2 & 2.2 & 1.0 & 5.2    \\

\hline
\end{tabular}
\end{table}  

  We have normalised spectra of  the catalogue as well as our program stars 
 by  dividing it  by   the continuum.  
 On these normalised spectra we measured the strength of several
 features (Table 2) that are known to be good  
 indicators of  spectral type  (and hence temperature), and luminosity class.

 Hydrogen line strengths,  
   particularly the  higher members of Balmer series show  different locii  
   for different luminosity classes in line strength versus (B-V) plot. 
 Figure 1 shows the variations of these features as a  function of (B-V)$_{o}$.
   We did not use Ca II K feature as this feature gets contaminated by
   H$_{\epsilon}$ for A type stars.
   Figure 1  also shows the polynomial fit for the strengths of hydrogen
   lines. The same curves (polynomial fit to the catalogue stars of Jacoby
   et al. 1984) are retained in Figure 2 but the data points are  of
   the EC stars observed by us.
\begin{figure}
\psfig{figure=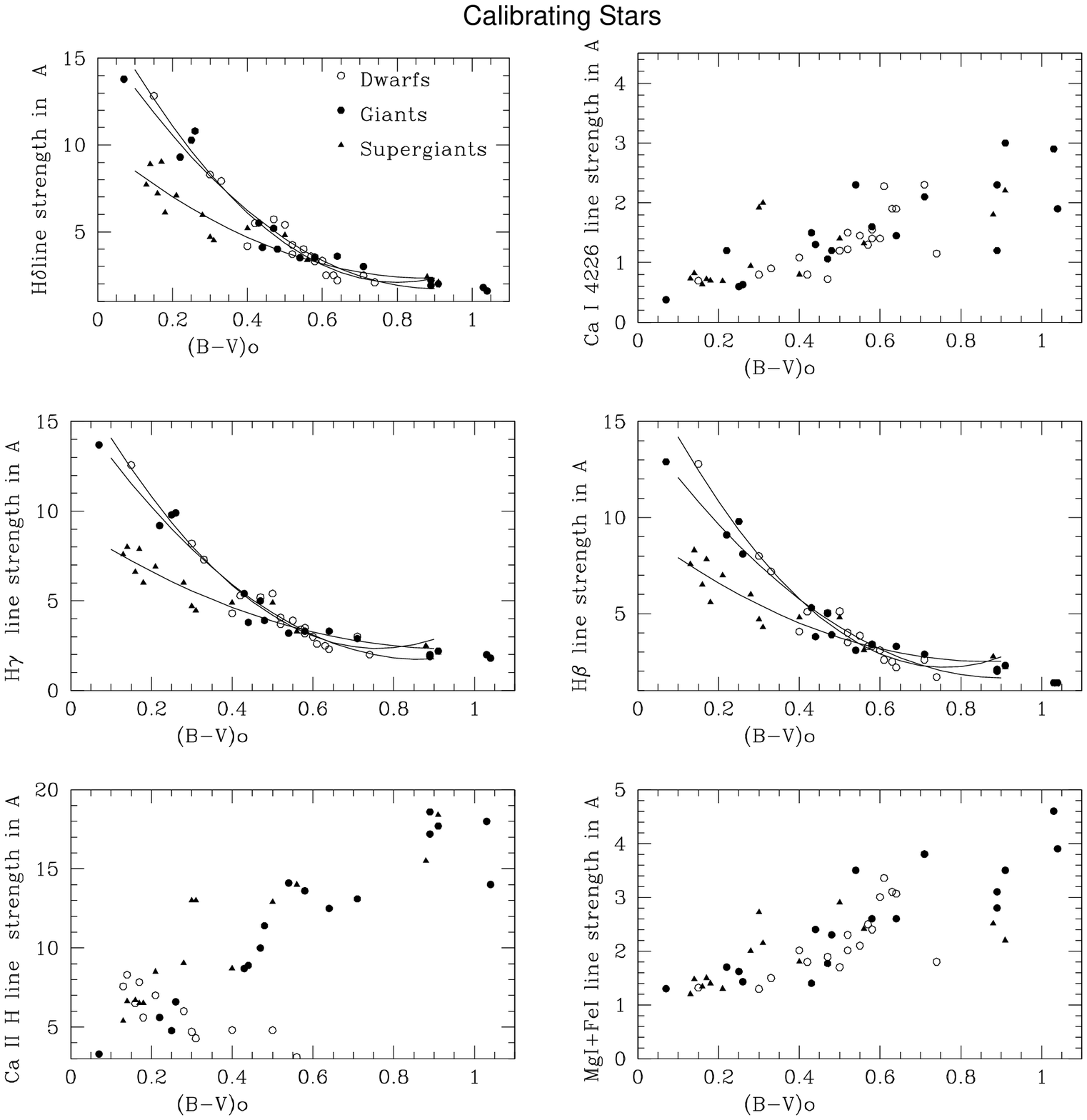,width=\hsize,height=20cm}
\vskip 0.2cm
 Fig.1 : Strengths of different spetral features as a function 
of (B-V)$_{o}$ for the stars from the Catalogue of Jacoby  at al. 
The symbols for all the six 
 panels have the same meaning as given in the panel 1.
\end{figure}

\begin{figure}
\psfig{figure=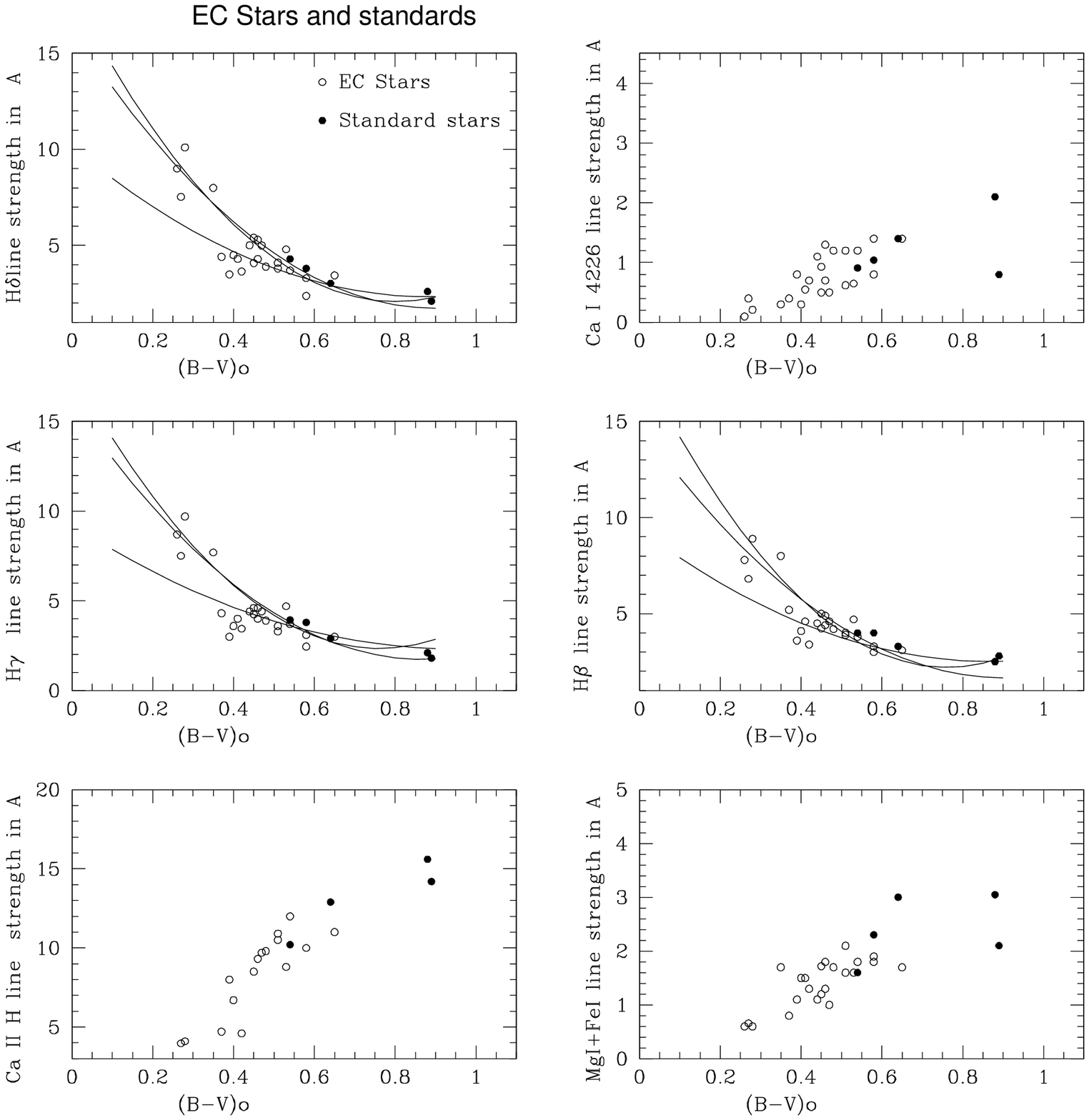,width=\hsize,height=20cm}
\vskip 0.2cm
Fig. 2 :  Strengths of different spetral features as a function of (B-V)$_{o}$ for
 the stars  observed by us with VBT. The symbols for all the six 
 panels have the same meaning as given in the panel 1.
\end{figure}

Strengths of the observed features for the program stars and the standard
 stars observed with OMR spectrograph are 
presented  in Table 3. The features 
   Ca II H and K lines, Ca I line at 4226\AA~  and MgI+FeI blend
  at 5170\AA~ are strongly dependent on temperature and hence show
  steep variations when plotted as a function of (B-V)$_{o}$. 
   Ca features show a linear relation with (B-V) with  no significant 
  distinction between luminosity types.  The Mg I features in 5167-5184 $\AA$~
 region show strong luminosity dependence at temperatures cooler than 6500K.
\begin{table}
{\bf Table 3: Strengths of observed features in \AA~ for the program stars and standards}\\
\begin{tabular}{ccccccccccc}
         &      &          &      &      &     &     &     &     &     &    \\
\hline
         &      &      Ca II K & Ca II H & $H_{\delta}$ & Ca I & $H_{\gamma}$& $H_{\beta}$&Mg+Fe & Mg+Fe & Fe I  \\
{\bf Star }&$(B-V)_{o}$& 3934& 3969& 4102& 4227 & 4340& 4861& 5171& 5184& 5270 \\ 
\hline
EC 05148-2731 & 0.28 &  4.1 &  --- & 10.1 & 0.2 & 9.7 &  8.9 & 0.6 & 0.2 & 0.5    \\
EC 10488-1244 & 0.45 &  8.5 &  8.4 &  5.4 & 0.5 & 4.6 &  5.0 & 1.2 & 0.8 & 1.5    \\
EC 10004-1405 & 0.40 &  6.7 &  --- &  4.5 & 0.3 & 3.6 &  4.1 & 1.5 & 0.6 & 1.4    \\
EC 10292-0956 & 0.54 & 12.0 & 10.0 &  3.7 & 1.2 & 3.7 &  3.8 & 1.8 & 1.1 & 1.7    \\
EC 11091-3239 & 0.47 &  9.7 &  8.4 &  5.0 & 0.5 & 4.4 &  4.6 & 1.0 & 0.4 & 1.5    \\
EC 11175-3214 & 0.35 &  --- &  --- &  8.0 & 0.3 & 7.7 &  8.0 & 1.7 & 0.7 & 1.9    \\
EC 11553-2731 & 0.41 &  --- &  --- &  4.3 & 0.6 & 4.0 &  4.6 & 1.5 & 0.7 & 1.4    \\
EC 11260-2413 & 0.26 &  1.7 &  --- &  9.0 & 0.1 & 8.7 &  7.8 & 0.6 & 0.1 & 0.3    \\
EC 12245-2211 & 0.44 &  --- &  --- &  5.0 & 1.1 & 4.4 &  4.5 & 1.1 & --- & 1.3    \\
EC 12418-3240 & 0.58 & 10.0 &  9.0 &  3.3 & 1.4 & 3.1 &  3.3 & 1.8 & 0.7 & 2.1    \\
HD 86801      & 0.58 &  --- &  --- &  3.8 & 1.0 & 3.8 &  4.0 & 2.3 & 1.3 & 2.2    \\
HR 5694       & 0.54 & 10.2 &  9.3 &  4.3 & 0.9 & 3.9 &  4.0 & 1.6 & 1.0 & 1.7    \\
EC 12473-1945 & 0.51 & 10.5 &  9.3 &  3.8 & 0.6 & 3.6 &  4.0 & 1.6 & 0.6 & 1.8    \\
EC 12477-1711 & 0.53 &  8.8 &  8.3 &  4.8 & 0.7 & 4.7 &  4.7 & 1.6 & 0.8 & 1.7    \\
EC 12477-1724 & 0.46 &  --- &  --- &  5.3 & 0.7 & 4.6 &  4.9 & 1.8 & 1.4 & 2.1    \\
EC 12493-2149 & 0.58 &  --- &  --- &  2.4 & 0.8 & 2.5 &  3.0 & 1.9 & 1.0 & 2.4    \\
EC 13042-2740 & 0.45 &  --- &  --- &  4.1 & 0.9 & 4.3 &  4.2 & 1.7 & 1.4 & 1.8    \\
EC 13501-1758 & 0.65 & 11.0 &  9.3 &  3.5 & 1.4 & 3.0 &  3.1 & 1.7 & 1.0 & 2.4    \\
HR 5384       & 0.64 & 12.9 & 10.0 &  3.0 & 1.4 & 2.9 &  3.3 & 3.0 & 1.8 & 2.6    \\
EC 13506-1845 & 0.48 &  9.8 &  --- &  3.9 & 1.2 & 3.9 &  4.2 & 1.7 & 1.0 & 2.1    \\
EC 13564-2249 & 0.51 & 10.9 &  9.8 &  4.1 & 1.2 & 3.3 &  3.9 & 2.1 & 1.2 & 1.5    \\
EC 13567-2235 & 0.46 &  9.3 &  8.4 &  4.3 & 1.3 & 4.0 &  4.4 & 1.3 & 0.6 & 1.0    \\
EC 13586-2220 & 0.37 &  4.7 &  6.4 &  4.4 & 0.4 & 4.3 &  5.2 & 0.8 & 0.6 & 1.1    \\
EC 14005-2224 & 0.42 &  4.6 &  7.0 &  3.7 & 0.7 & 3.5 &  3.4 & 1.3 & 0.4 & 0.7    \\
EC 15473-1241 & 0.27 &  4.0 &  8.0 &  7.5 & 0.4 & 7.5 &  6.8 & 0.7 & 0.1 & 0.4    \\
HR 4786       & 0.89 & 14.2 & 12.3 &  2.1 & 0.8 & 1.8 &  2.8 & 2.1 & 1.0 & 3.0    \\

\hline
\end{tabular}
\end{table}

   Using the  hydrogen lines strengths, the first estimates of the spectral types and 
   luminosity classes were made   for the program stars. 
  The standard stars observed were  also
    analysed in an   identical way. The comparison of the results obtained
   by us with the published ones for the standard stars lead us to believe that
   the spectral types are accurate to $\pm$ 2 subtypes. 
   Since the  bulk of classifying stars  belong to luminosity classes  V, 
III and I with a very few of type II, the errors are difficult  to estimate for 
luminosity types. 
   We have used  luminosity types determined above, the (B-V)$_{o}$
  from Beers et al.(2001)  and the calibration of Schmidt-
   Kaler (1982) to derive  T$_{eff}$ of the program stars.
  Our estimates of T$_{eff}$ and log g (estimated from the luminosity types)
  that are based on the hydrogen line strengths are presented  in columns
  4 and 6 respectively of Table 4 under the subheading H.
\begin{table}
{\bf Table 4: Derived Atmospheric Parameters and Metallicities  } \\
\begin{tabular}{cclccccccccc}
          &       &&     &&   &       &      &     &    &    &   \\
\hline
{\bf Star} &{\bf Sp Type } & LC & $T_{eff}$ & $T_{eff}$  &log~g & log~g  &
$(B-V)_{o}$& M$_{v}$ & Dist. &  [Fe/H] & Z \\
           &      &      &   H    &   S  &  H  &   S  & & & kpc    & S & kpc  \\ 
\hline
EC 05148-2731 & F0 & IV  &  7500 & 8250 &3.6 & 4.0  & 0.28 & 1.7 & 1.58 & -0.6 & 0.76 \\ 
EC 10488-1244 & F5 & III &  6500 & 6500 &3.0 & 2.75 & 0.45 & 1.6 & 0.79 & -1.0 & 0.47  \\
EC 10004-1405 & F3 & IV  &  6500 & 7300 &2.5 & 4.0  & 0.40 & 2.5 & 1.69 & -0.7 & 0.81 \\
EC 10292-0956 & F9 & IV  &  6100 & 6000 &3.5 & 4.25 & 0.54 & 2.8 & 0.71 & -1.2 & 0.41 \\ 
EC 11091-3239 & F6 & IV  &  6400 & 6500 &3.0 & 4.25 & 0.47 & 2.7 & 0.93 &  0.0 & 0.36 \\
EC 11175-3214 & F2 & IV  &  6900 & 7300 &4.0 & 3.5  & 0.35 & 2.0 & 1.26 & -0.5 & 0.51 \\
EC 11553-2731 & F3 & IV  &  6500 & 6800 &3.0 & 4.0  & 0.41 & 2.5 & 0.57 & -0.5 & 0.29 \\
EC 12245-2211 & F5 & III &  6500 & 6800 &3.0 & 3.0  & 0.44 & 1.6 & 2.96 & -1.0 & 1.74 \\
EC 12418-3240 & G0 & IV  &  6100 & 6000 &4.0 & 4.0  & 0.58 & 3.0 & 0.79 & -0.9 & 0.35 \\
EC 12473-1945 & F8 & III &  6200 & 6300 &3.5 & 3.75 & 0.51 & 1.6 & 0.58 & -0.9 & 0.36 \\ 
EC 12477-1711 & F8 & III  &  6200 & 6300 &3.5 & 3.75 & 0.53 & 1.8 & 0.95 & -0.8 & 0.62 \\ 
EC 12477-1724 & F6 & IV  &  6400 & 6500 &4.0 & 4.25 & 0.46 &  2.6 & 0.57 & -0.5 & 0.36 \\
EC 12493-2149 & G0 &  V  &  6000 & 6000 &4.5 & 5.0  & 0.58 & 4.3 & 0.69 & -0.7 & 0.41 \\
EC 13042-2740 & F6 & III  & 6300 & 6300 &2.0 & 3.0  & 0.45 & 1.5 & 1.95 & -1.0 & 1.01  \\
EC 13501-1758 & G0 & IV  &  5800 & 5500 &3.0 & 4.0  & 0.65 & 3.0 & 0.84 & -1.3 & 0.52 \\
EC 13506-1845 & F6 & III &  6300 & 6300 &3.0 & 3.0  & 0.48 & 1.6 & 1.79 & -0.7 & 1.08 \\
EC 13564-2249 & F7 & III &  6200 & 6000 &2.0 & 3.5  & 0.51 & 2.0 & 1.43 & -0.9 & 0.78 \\
EC 13567-2235 & F6 & III &  6300 & 6500 &2.0 & 3.0  & 0.46 & 1.6 & 1.58 & -1.0 & 0.87 \\ 
EC 13586-2220 & F3 & III &  6700 & 6800 &2.5 & 3.0  & 0.37 & 1.6 & 2.62 &  -1.8 & 1.45 \\
          &       &     &&   &       &      &     &    & &  &   \\
 Standard Stars    &       &     &&   &       &      &  &  &    &    &   \\
          &       &     &&   &       &      &     &    &&   &   \\
\hline
HR 4054       & F6 & IV  &  6300 & 6300 &3.9 & 4.0  & 0.45 & 2.6 &  0.026 & 0.0 & 0.01 \\
HD 86801      & G0 &  V  &  6050 & 5800 &4.5 & 4.75 & 0.58 & 4.3 & 0.079 & -1.2 & 0.06 \\ 
HR 5694       & F8 & IV  &  6200 & 6300 &4.0 & 4.0  & 0.54 & 2.9 & 0.027 & -0.2 & 0.02 \\   
HR 5384       & G1 &  V  &  5800 & 5800 &4.5 & 4.25 & 0.64 & 4.9 & 0.019 & -0.4 & 0.01 \\
\hline
\end{tabular}
\end{table}

\section{Spectrum Synthesis}
 Starting with these
 values  of T$_{eff}$  and log g           derived in the last
 section  we decided to further refine these parameters and also
 to derive metallicity by using line and spectrun synthesis method.

\subsection{  Sensitivity of Mg I features at 5167-5184 $\AA$ }

 The three Mg I features in 5167-5184\AA\ region were found to be
 very sensitive to temperature and gravity changes  but show stronger dependence
 on gravity for stars cooler than 6500K. We  have computed the line strengths of
 these features using  2000 version of spectrum synthesis program written by Sneden 
 (first version described in Sneden 1973) and  using models of Kurucz (1993)
   covering a large range in T$_{eff}$ and log g. We show in the figure 3
   computed blend equivalent widths (sum of equivalent widths for the three Mg I lines )
   of  Mg I features in 5167-5184 $\AA$~ for models of different temperatures 
 and gravities. This figure was found very useful in the refinement of temperature and
 gravity  estimates particularly the gravity.

\subsection{ Final atmospheric parameters and estimation of [Fe/H]}
 We have synthesized the spectral regions around 4900-5400\AA~ 
 and redetermined more accurately, the stellar parameters T$_{eff}$ and log~g
as well as metallicity [Fe/H] for the  program stars. We were guided by Figure
3 in improving the gravity estimates. 
\begin{figure}
\psfig{figure=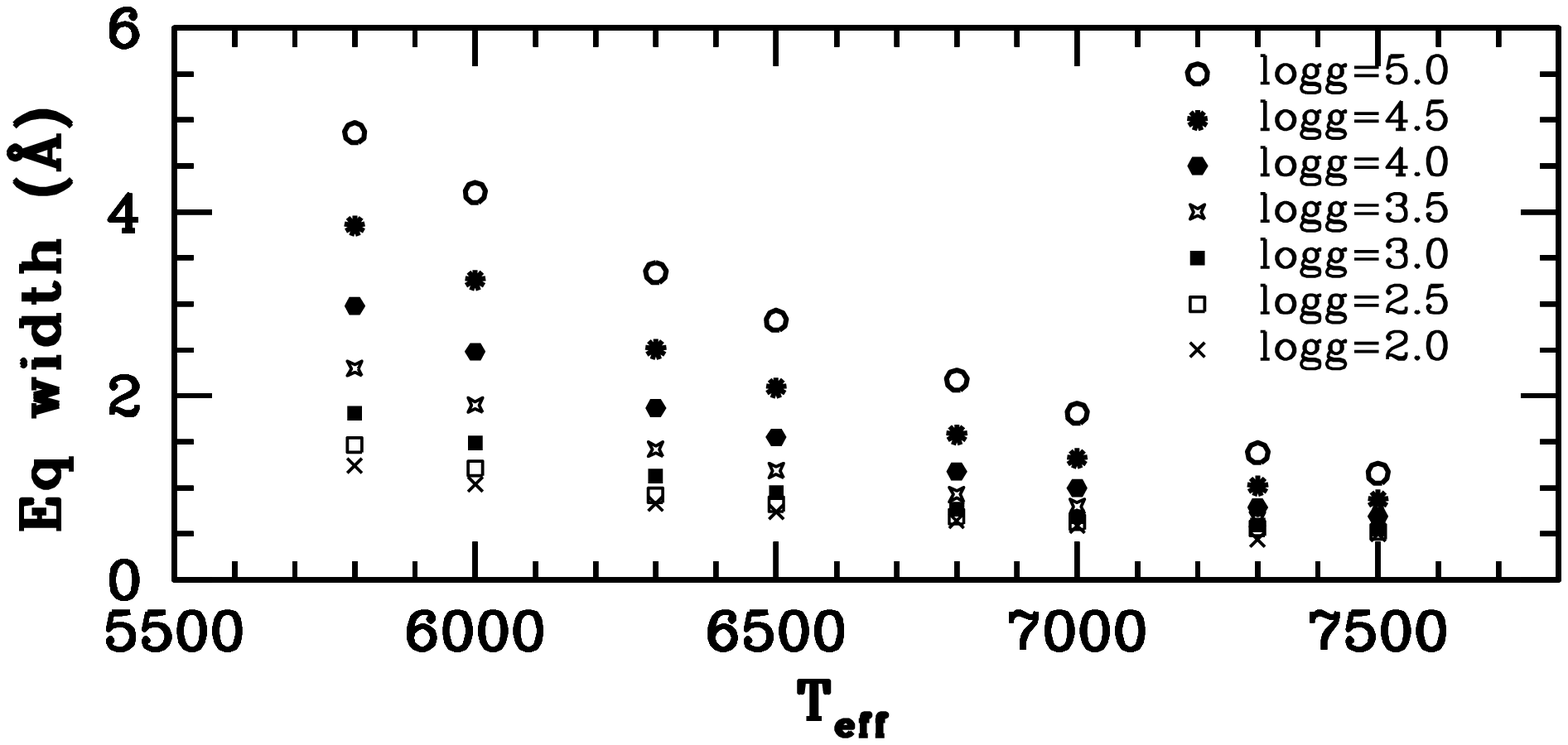,width=\hsize,height=20.cm}
 Fig. 3 : Computed equivalent widths (in \AA~) for Mg I feature at 5174-5183 \AA~
         for different temperatures and gravities. Each point corresponds to
        combined equivalent width of three Mg I lines that form  this feature.
\end{figure}
 For synthesizing the spectral regions we have generated a linelist using
 NIST data base,  supplemented by linelists 
 of Thevenin (1989, 1990),  unpublished compilation of R.E. Luck and  
 linelist database of Kurucz.
For checking the  accuracy of atomic data, mainly the
oscillator strengths used in the linelist, we had synthesized the solar 
spectrum around the spectral region of our interest using the solar model
of Holweger and Muller (1974).
A  fairly good reproduction of the solar spectrum obtained using this linelist
lent support to our belief that the linelist was satisfactory. 
 We have further tested the linelist by computing the above mentioned 
spectral regions for two
standard stars, HR 5384 and HR 5694,  of known  stellar parameters.  
 A temperature of T$_{eff}$
$\sim$ 5640K and metallicity [Fe/H] = -0.20 was estimated for HR 5384 by
Lebreton et al. (1999). Cayrel de Stobel et al. (1992) had reported  
temperature range  T$_{eff}$ $\sim$ 5950 to 6260 K, log~g $\sim$ 3.9 to 4.0 and
metallicity [Fe/H] $\sim$ -0.6 to 0.06 for HR 5694. A  synthetic spectrum 
obtained using the adopted linelist and  a stellar model with  T$_{eff}$ of 5800K,
 log~g = 4.25 and
metalicity [Fe/H] = -0.22 could  reproduce the observed spectrum of HR 5384
fairly in good agreement. Similarly,  a synthetic spectrum  obtained using
a  stellar model with  T$_{eff}$ of 6300 and log~g = 4.0 could also reproduce
the observed spectrum of HR 5694.  These agreements are  illustrated in 
figure  4 along with the synthetic spectra of two other standard stars,
HR4054 and HD86801,  all well-known  radial velocity standards. 
\begin{figure}
\psfig{figure=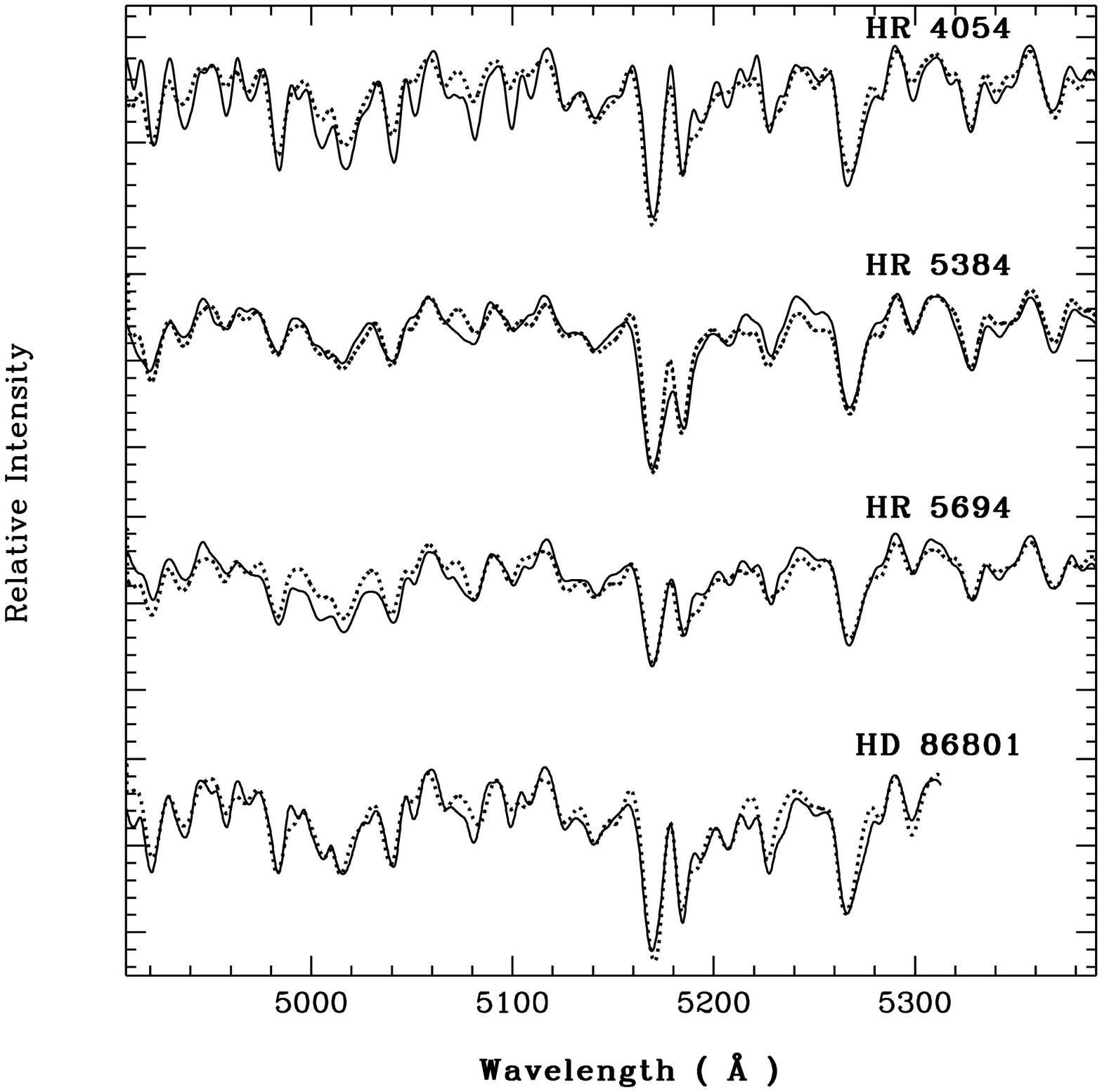,width=\hsize,height=20cm}
\vskip 0.2cm
 Fig. 4: The observed spectra (shown by continuous line) of standard stars
          compared with synthesized spectra.
\end{figure}

For spectral synthesis of the program stars we have started with
 atmospheric models of Gustafsson et al. (1975), and Kurucz (1993)
 at T$_{eff}$ and log~g  estimated from the hydrogen line strengths.
  The gravity estimates were further refined with the help of figure 3.

 Starting from solar metallicity  models, 
the value of [Fe/H] is gradually changed to obtain a best fit of the
observed spectrum. While changing [Fe/H], [X/H] of the elements which are
identified as   having maximum contributions in the spectral region of 
our interest , such as Mg, Ca, Ti, Ni are also correspondingly changed.
 For this, we have made use of [Fe/H] vs [X/H] compiled from various
 sources and compared with their evolutionary calculations by 
Goswami and Prantzos (2000, fig 7). The observed data has              
 large scatter  in {[Fe/H] range -0.5 to 0.0.  For our purpose of spectral
synthesis we have adopted an average value of [X/Fe] corresponding to [Fe/H].
In figures 5 and 6 we present synthetic spectra for several of our
program stars compared with their observed spectra. The results are 
presented in Table 4, where H and S denote  atmospheric parameters derived from 
hydrogen lines strengths and spectral synthesis respectively. 
\begin{figure}
\psfig{figure=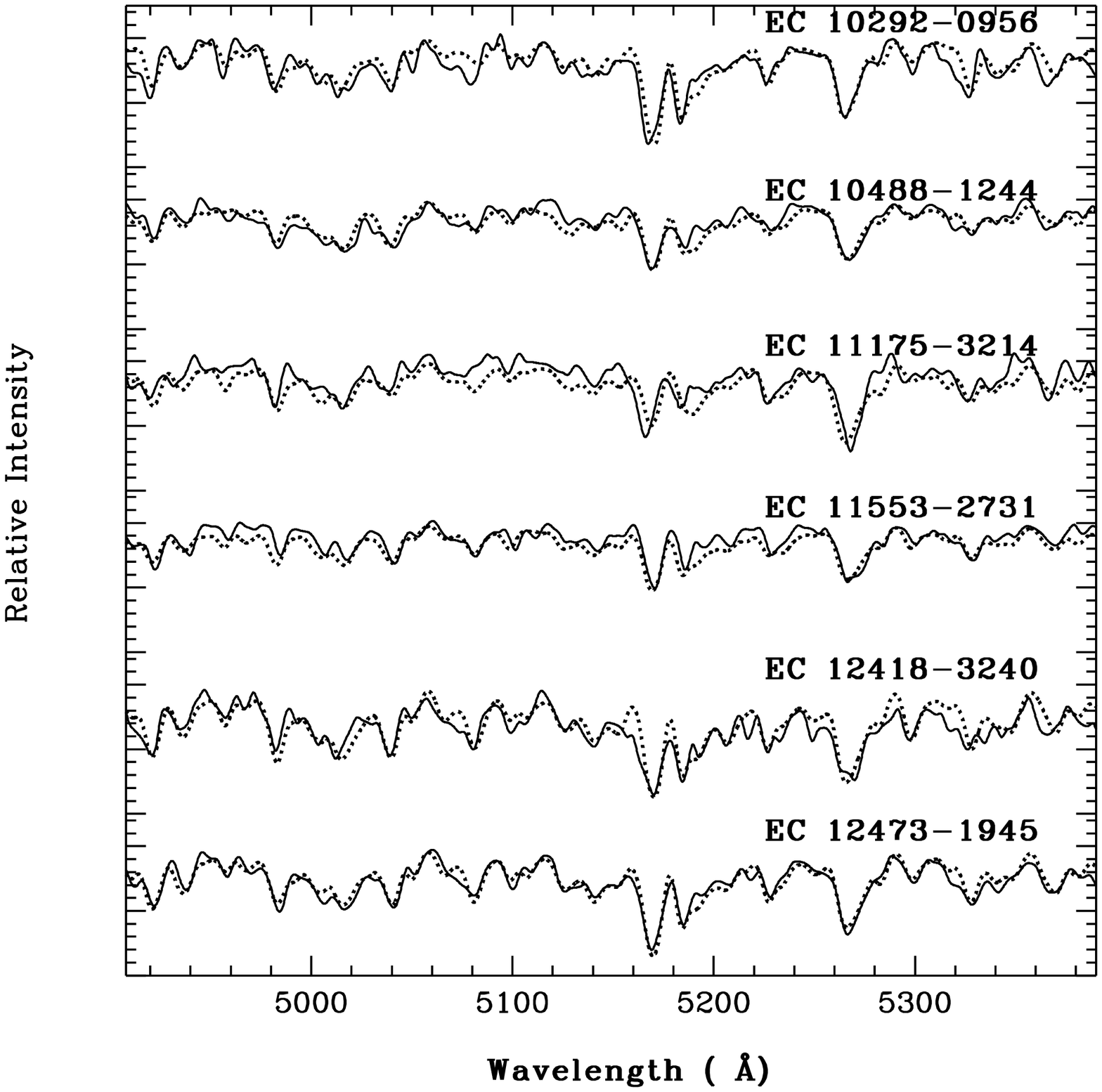,width=\hsize,height=20cm}
\vskip 0.2cm
 Figs. 5: The observed spectra (shown by continuous line) of program stars
          compared with synthesized spectra.
\end{figure}
  
\begin{figure}
\psfig{figure=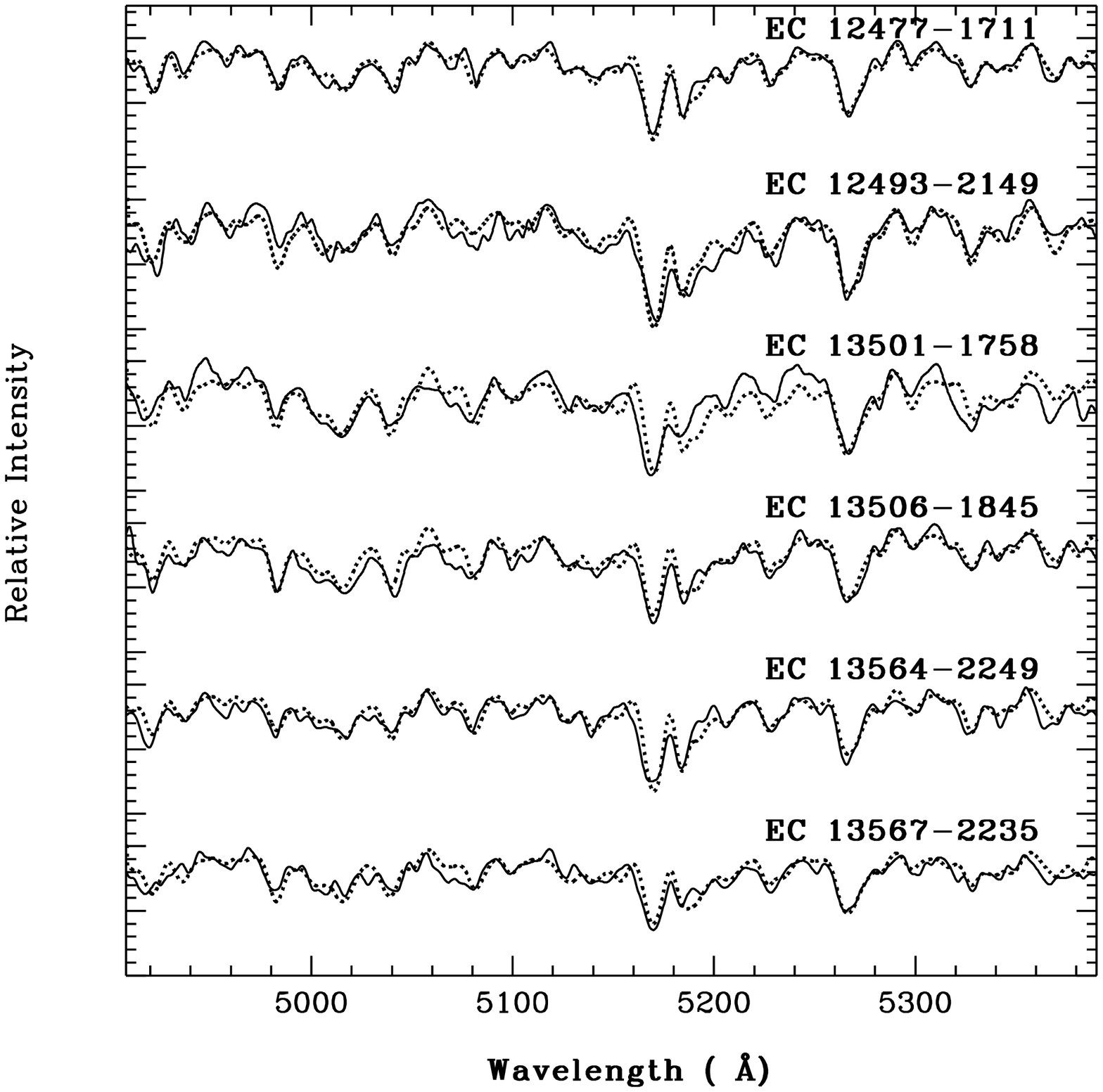,width=\hsize,height=20cm}
\vskip 0.2cm
 Figs. 6: The observed spectra (shown by continuous line) of program stars
          compared with synthesized spectra.
\end{figure}

 The agreement between the derived T$_{eff}$ and log ~g  from  the two approches
 is not very satisfactory for the stars hotter than
 7000K but is very good         for stars in 6000- 7000K temperature range 
 where most of the sample star belong to. 
 For  majority of stars T$_{eff}$ agrees within $\pm$ 200K. We seem to derive systematically
 higher value of gravity from spectrum synthesis but we are more inclined 
 to believe higher values
as Mg I feature is known to be a good gravity indicator as illustrated in
 Figure 3.
 Secondly, the calibrating stars in Jacoby's catalogue did not have uniform
 distribution in the luminosity classes. We believe our [Fe/H] estimates for
 stars in T$_{eff}$ 5800-7000K are accurate within $\pm$ 0.3 dex.

 Perhaps for hotter stars one requires non-LTE models and a linelist tested for
 completeness using some  well studied  hot star like Vega.
 It is likely that for EC 05148-2731 and EC 10004-1405 the estimated [Fe/H] may have
 uncertainty  larger than  $\pm$ 0.3 .

 From the spectroscopic estimates of T$_{eff}$ and log ~g and calibration of 
 Schmidt-Kaler we have estimated absolute magnitude and hence the distances 
 that are also presented in Table 4. 
 The last column of the Table 4 shows the distance above the galactic plane in kpc.
 The last four entries in the table are the radial velocity standards observed
 with  the same  setup. For HR 5384 and HR 5694 the atmospheric parameters and 
 [Fe/H] are in close agreement with the published values. No previous [Fe/H] 
 estimate exists for HD 86801.
 For EC 10488-1244,  EC 10004-1405 and EC 10292-0956, Prof. T.C. Beers had 
 sent  us unpublished [Fe/H] estimates of -1.1, -0.8 and -1.4 that are in very
 good agreement with values -1.0, -0.7 and -1.2 derived by us though his
 estimate  is based upon the strength of Ca II K line  . 

\section{Results and Conclusions}
  It is obvious from the Table 4 that the bulk of the stars chosen from the
  EC survey belong to spectral type range A9 to F7. A large fraction of
  them are dwarfs or subgaints and the rest are giants. 
  Sample stars    EC12477-1724, EC10004-1405, EC11260-2413   have
  radial velocities in excess of 100 kms$^{-1}$ (+230,+138,-100 kms$^{-1}$ )
  $\pm$ 15 kms$^{-1}$ respectively.
  However, they are not the most metal-poor objects. In fact the significantly 
  metal-poor stars like EC 13586-2220, EC 13501-1758 and EC 10292-0956
  are not the high velocity objects (radial velocity is less than 50 kms $^{-1}$) .
 Hence the high velocity is not a very robust criteria  for searching new metal-poor stars.
  From the inspection of galactic coordinates and  derived distances, it  
  appears that the sample stars are distributed in the thick disk around the  local
  spiral arm. The most metal-deficient star EC 13586-2220 with Galactic 
  longitude 323$^{o}$ and distance of 2.6 kpc, possibly lies in the
  direction of 
  interarm region between the local and Sagittarius arm. With a distance of
  1.45 kpc above the Galactic plane, it might be beyond the thick disk.
   Nearly half the sample stars covered in the present work turned out to be
   significantly metal-poor with [Fe/H] in -0.7 to -1.2 range.  EC 13586-2220, 
   EC 13501-1758 and EC 10292-0956 are the most metal-poor objects  in the sample
   with [Fe/H] of -1.8, -1.3 and -1.2 respectively. We intend to carry out
   a complete analysis of these three objects using high resolution spectra.
   Though the present sample is not very extensive, we do not see strong 
   dependence of metallicity on the Z.
   We conclude that the objects presented in EC survey are promising candidates
  for metal-poor stars and detailed investigation of a larger sample
   will lead to a better understanding of the galactic structure and
   that  of early Galactic history. The present work presents the initial 
   phase of an ongoing survey program.
\section*{Acknowledgements}

   We are grateful to Prof. T.E. Beers for sending us the list of candidate stars
   and his unpublished [Fe/H].  We also thank Prof. D. L. Lambert for  
   useful comments on the manuscript. We are grateful to
   VBT staff members for their kind help in getting observations and
   also  to an anonymous referee whose comments have led to a 
    much better presentation of the material in this paper.                      

{  }
 
\end{document}